\documentclass[onecolumn,prc]{revtex4-2}
\usepackage{graphicx}
\usepackage[caption=false]{subfig}

\newcommand{\beq}{\begin{eqnarray}}
\newcommand{\eeq}{\end{eqnarray}}

\begin{document}


\title{Dilepton Correlations from Heavy Flavor Decays}

\author{T. Dahms\footnote{Now at DXC Technology, Munich, Germany.}}
 \affiliation{Physics Department and Excellence Cluster `Universe', Technical University Munich, 85748 Garching, Germany.}
\author{R. Vogt}%
 \affiliation{Nuclear and Chemical Sciences Division,
  Lawrence Livermore National Laboratory, Livermore, CA 94551, USA}
 \affiliation{Physics and Astronomy Department, University of California, Davis, CA 95616, USA.
}%

\date{\today}

\begin{abstract}
  {\bf Background:} Azimuthal correlations between heavy flavor hadrons have been previously studied in $p+p$ collisions (Phys. Rev. C {\bf 98}, 034907 (2018), {\bf 101}, 034910 (2020)).  These studies found good agreement with the data and provide a baseline for further studies in $p+A$ and $A+A$ collisions.  
  {\bf Purpose:} This work extends those studies to heavy flavor hadron decays to low mass lepton pairs.  The low mass dilepton region is important for heavy ion collisions because of the interest in thermal dilepton production as a signature of the early time dynamics of the medium.
  {\bf Methods:} Building on previous work, azimuthal correlations between leptons from semileptonic decays of heavy flavor hadrons are examined.  The exclusive \textsc{HVQMNR} code is used in the calculations, made in the PHENIX acceptance at $\sqrt{s} = 200$~GeV and for dielectrons in ALICE at $\sqrt{s} = 13$~TeV.  The $b \overline b$ decay contributions subtract like-sign lepton pairs from the opposite-sign signal.
  {\bf Results:}  The next-to-leading order calculations reproduce the trends of the PHENIX data.  The calculations at 13~TeV show a change in signal as electron pair $p_T$ is increased, with the peak of the $\Delta \phi$ distribution moving from $\Delta \phi \sim \pi$ to $\sim 0$ and the contribution from bottom decays becoming dominant.  The sensitivity of the signal to $k_T$ broadening is also studied and found to be small.
  {\bf Conclusions:}  It is found that the dependence on $k_T$ broadening previously observed is significantly reduced by the decay process. Despite this, the correlations between decay leptons retains some memory of the correlations between the parent hadrons in $p+p$ collisions.  However, to study these correlations in heavy-ion collisions, it is necessary to separate them from thermal dilepton production in the same mass region.
\end{abstract}

\maketitle


\section{Introduction}
\label{intro}

Measurements of low-mass lepton pairs (dileptons, $m_{ll}$) in relativistic heavy-ion collisions have historically been proposed as probes of thermal radiation from the quark-gluon plasma (QGP) and of possible mass modifications of short-lived vector mesons due to chiral symmetry restoration, see reviews in Refs.~\cite{David:2006sr,Rapp:2013nxa,Geurts:2022xmk}.  Thermal dileptons are a promptt signature of the early time dynamics of the QGP, with higher mass dileptons, $2 < m_{ll} < 4$~GeV, produced in the first $1-2$~fm/$c$ \cite{Massen:2024pnj}.  However, in the  mass region above the light quark resonances ($\rho$, $\omega$ and $\phi$) and below the $J/\psi$ mass, $1 < m_{ll} < 3$ GeV, heavy quark contributions are dominant.  Different attributes of charm and bottom production can help separate the two contributions and provide an additional handle on their production characteristics.

The center of mass energies in heavy-ion collisions have increased by more than two orders of magnitude over the last few decades, from $\sqrt{s_{NN}} \sim 20$~GeV at the SPS to 200~GeV at RHIC and 5.02~TeV at the LHC. This rise has affected the physics available from dilepton measurements.  The increased energy has led to higher temperatures in heavy-ion collisions, implying a stronger contribution from thermal dilepton production.  However, it has also led to an even stronger increase of the contribution of hard probes to the dilepton continuum.  Drell-Yan dileptons can dominate the high invariant mass region.  At LHC energies, $Z^0 \rightarrow l^+ l^-$ measurements have been made in $p+{\rm Pb}$ and Pb+Pb collisions \cite{CMS:2021ynu,ALICE:2020jff,ALICE:2016rzo}.  The quark and antiquark distributions that dominate the $Z^0$ production process, $q \overline q \rightarrow Z^0 X$, resulting in high mass, large $Q^2$ vector bosons in the range $0.0003 < x < 0.4$ \cite{Eskola:2021nhw}, are subject to modifications in the cold nuclear medium \cite{Vogt:2000hp}.  The $Z^0$ measurements in $p+{\rm Pb}$ have been used in global analyses of the modifications of sea quark distributions in nuclei \cite{Eskola:2021nhw}.  Charm and bottom decays contribute to the dilepton continuum through simultaneous semileptonic decays of correlated heavy flavor hadrons, such as $c \overline c \rightarrow D^+ D^- \rightarrow e^+ e^- {\rm X}$.  The contributions from these non-prompt correlated decays increases with $\sqrt{s_{NN}}$.  The large number of $c \overline c$ pairs produced in central heavy-ion collisions in particular results in a considerable contribution to dilepton production from uncorrelated $Q \overline Q$ pair decays \cite{Vogt:1993kn,Gavin:1996bx,Kumar:2012qx}.  While these pairs can be removed by like-sign subtraction, there is an irreducible contribution to the dilepton continuum from these days.  In the mass region below the $\Upsilon$, these decays dominate over Drell-Yan dileptons.

All these contributions to the mass distribution cannot be separated, even in a given event, but must be modeled.  However, other kinematic characteristics can be employed to distinguish between the dilepton sources.  The azimuthal separation of thermal dileptons is expected to be isotropic.  On the other hand, the azimuthal $Q \overline Q$ correlations are not isotropic with decay characteristics that depend on their production process. However, it is unclear whether the production characteristics of the heavy quark pairs survive their semileptonic decays.  In its rest frame, a heavy hadron decays such that its decay products are distributed isotropically.  However, when transforming back to the laboratory (measurement) frame, the decay products will be boosted in the direction of their decay parent so they should, at least to some extent, retain the memory of the azimuthal correlations of the parent $Q \overline Q$ pair.  Therefore, it is important to study their production and decay characteristics in $p+p$ collisions as a measurement baseline.  An advantage of studying correlated heavy flavor decays to dileptons in $p+p$ collisions is their low multiplicity in such events.  Since only a single $Q \overline Q$ pair is most likely to be produced in these collisions, aside from ``ultra-central'' collisions with high final-state multiplicities \cite{ATLAS:2015hkr}, any $Q \overline Q$ pair production should be correlated. Heavy flavor correlations in $p+p$ collisions were studied in Refs.~\cite{Vogt:2018oje,Vogt:2019xmm} with particular attention to how fragmentation, $k_T$ broadening, and the $p_T$ range of the heavy flavor measurement could affect the results.  That work is followed up here by extending the calculation to heavy flavor semileptonic decays to leptons.

The model calculation is described in Sec.~\ref{sec:model}.  A brief description of how the lepton decays are handled in the calculation is given in Sec.~\ref{sec:decays}.   Calculations for low mass dileptons are presented in Sec.~\ref{sec:calcs_for_expts} for $p+p$ collisions at $\sqrt{s} = 200$~GeV and 13~TeV.  Results are presented for the azimuthal separation between the leptons as well as the lepton pair mass distributions.  The sensitivity of the results to the introduced $k_T$ broadening is given in Sec.~\ref{sec:Delta_kt}.  A summary is given at the end.

\section{Model Description}
\label{sec:model}

The calculations here follow those outlined in Ref.~\cite{Vogt:2018oje}.
The \textsc{HVQMNR} code \cite{Mangano:1991jk}, also used here, was designed to
calculate $Q \overline Q$ pair production at NLO.
It is necessary to use a code such as the exclusive \textsc{HVQMNR} calculation
because the $Q \overline Q$ pair quantities are calculable in such an approach
while only single inclusive quark distributions are available with the FONLL
\cite{Cacciari:1998it} and generalized mass, variable flavor number 
\cite{Kniehl:2004fy,Helenius:2018uul} approaches.

Hadronization of the heavy quarks to the final-state mesons
decaying to leptons was carried out by employing 
the Peterson fragmentation function, as in the \textsc{HVQMNR} code.  
The calculation also includes $k_T$ broadening which, while not part of the
hadronization process, was incorporated into the \textsc{HVQMNR} code to offset
the strong modification of the charm momentum distribution resulting from the
default Peterson fragmentation function \cite{MLM1}.

Intrinsic $k_T$ broadening is
related to QCD resummation at low $p_T$ \cite{CYLO}.  It was applied first to
the initial $q$ and $\overline q$ distributions in Drell-Yan production, see
{\it e.g.} Refs.~\cite{CYLO,APP,CGreco}.  In the
\textsc{HVQMNR} code, the intrinsic $k_T$ is added in the final state, rather than in
the initial state.  The Gaussian distribution, $g_p(k_T)$ \cite{MLM1},
\begin{eqnarray}
g_p(k_T) = \frac{1}{\pi \langle k_T^2 \rangle} \exp(-k_T^2/\langle k_T^2
\rangle) \, \, ,
\label{intkt}
\end{eqnarray}
multiplies the parton
distribution functions, 
assuming the $x$ and $k_T$ dependencies in the initial partons 
factorize.

The average broadening, $\langle k_T^2 \rangle$, in Eq.~(\ref{intkt})  was
fixed by comparing the measured quarkonium $p_T$ distributions to
$J/\psi$ and $\Upsilon$ production calculated in the color evaporation
model.  This model was implemented in the \textsc{HVQMNR} code by applying a cut on
the $Q \overline Q$ pair mass
\cite{Gavai:1994in}.  The effect of broadening decreases with increasing $\sqrt{s}$
energy since the average $p_T$ of the heavy quark distribution  increases,
decreasing the effect of the broadening at higher energies.
The energy dependence of $\langle k_T^2 \rangle$ in Ref.~\cite{Nelson:2012bc} is
\begin{eqnarray}
  \langle k_T^2 \rangle = 1 + \frac{\Delta}{n} \ln \left(\frac{\sqrt{s}}{20 \,
    {\rm GeV}} \right) \, \, {\rm GeV}^2 \, \, 
\label{eq:avekt}
\end{eqnarray}
with $\Delta = 1$.  The value of $n$ depends on the heavy quark mass with
$n = 12$ for charm and 3 for bottom.  The parameter $\Delta$ was introduced
in Ref.~\cite{Vogt:2018oje} to study the effect of $k_T$ broadening on the
azimuthal separation between the $Q$ and $\overline Q$.

The value of $\epsilon_P$, the Peterson function
parameter, was set by comparison to the FONLL $D$ and $B$ $p_T$ distributions
in Ref.~\cite{Vogt:2018oje}.  The FONLL $D$ meson fragmentation functions are
based on Mellin moments, including pseudoscalar and vector parts for $D^0$
production with the contributions for direct $c \rightarrow D^0$ fragmentation
and production through the excited state,
$c \rightarrow D^{*0}/D^{*+} \rightarrow D^0$.  In this case,
$\epsilon_c = 0.008$ gave the best agreement with the LHC $D$ meson data. On
the other hand, the bottom quark fragmentation function in FONLL is simply
$z(1-z)^{\epsilon_b}$.  Less of the bottom quark energy is lost in the
hadronization process, {\it i.e.}, $\langle z_b \rangle > \langle z_c \rangle$.
Thus, for $B$ meson
hadronization, $\epsilon_b = 0.0004$ is employed in the Peterson function.

The heavy quark mass, $m_Q$,
factorization scale, $\mu_F$, and renormalization scale $\mu_R$ and their uncertainties were set by comparison
to the $Q \overline Q$ total cross section data \cite{Vogt:2018oje}.  The mass
and scale parameters for charm quarks was found to be $(m_c,\mu_F/m_T, \mu_R/m_T) = (1.27 \pm 0.09~{\rm GeV}, 2.1^{+2.55}_{-0.85}, 1.6^{+0.11}_{-0.12})$
\cite{Nelson:2012bc} while, for bottom quarks,
$(m_b,\mu_F/m_T,\mu_R/m_T) = 4.65 \pm 0.09~{\rm GeV}, 1.40^{+0.77}_{-0.49},1.10^{+0.22}_{-0.20})$ was employed.

The mass and scale uncertainties on the distributions are 
calculated from the one standard deviation uncertainties on
the quark mass and scale parameters.  The mass uncertainty comes from using the
upper and lower limits of the heavy quark mass.
If the central, upper and lower limits
of $\mu_{R,F}/m_T$ are denoted as $C$, $H$, and $L$ respectively, seven
sets then determine the scale uncertainty: $\{(\mu_F/m_T,\mu_F/m_T)\}$ =
$\{$$(C,C)$, $(H,H)$, $(L,L)$, $(C,L)$, $(L,C)$, $(C,H)$, $(H,C)$$\}$.    
The uncertainty bands are then obtained
\cite{Nelson:2012bc,NVFinprep} by
adding the uncertainties from the mass and scale variations in 
quadrature. The band contained by the resulting curves,
\begin{eqnarray}
\frac{d\sigma_{\rm max}}{dX} & = & \frac{d\sigma_{\rm cent}}{dX} 
+ \sqrt{\left(\frac{d\sigma_{\mu ,{\rm max}}}{dX} -
  \frac{d\sigma_{\rm cent}}{dX}\right)^2
  + \left(\frac{d\sigma_{m, {\rm max}}}{dX} -
  \frac{d\sigma_{\rm cent}}{dX}\right)^2} \, \, , \label{sigmax} 
\\
\frac{d\sigma_{\rm min}}{dX} & = & \frac{d\sigma_{\rm cent}}{dX} 
- \sqrt{\left(\frac{d\sigma_{\mu ,{\rm min}}}{dX} -
  \frac{d\sigma_{\rm cent}}{dX}\right)^2
  + \left(\frac{d\sigma_{m, {\rm min}}}{dX} -
  \frac{d\sigma_{\rm cent}}{dX}\right)^2} \, \, , \label{sigmin} 
\end{eqnarray}
defines the uncertainty on the cross section.
Here $X$ denotes any kinematic observable. The cross section labeled ``cent''
denotes the calculation with the best fit values for
$m$, $\mu_F$ and $\mu_R$.  In the cross sections labeled $\mu$,
the mass is kept fixed to the central value while the scales are varied as
above.  The cross sections
labeled $m$ vary the mass while leaving the scales fixed at the central values.

At leading order, the $Q \overline Q$ pairs are produced back-to-back with
azimuthal separation $\phi = \pi$, a delta function.  At next-to-leading order,
both real and virtual corrections contribute.  Virtual corrections are soft
gluon exchanges that can decorrelate the LO delta function peak.  The real
$2 \rightarrow 3$ corrections on the other hand can give rise to the
possibility that a $Q \overline Q$ pair is produced against a hard parton in
the final state, leading to a secondary peak for high heavy quark $p_T$.
Integrating over all $p_T$, the distribution has a peak at $\phi = \pi$ and a
long tail as $\phi \rightarrow 0$.  Because the charm quarks are
lighter, this tail is enhanced relative to that of $b$ quarks.  Introducing a
fragmentation function has a small effect on the azimuthal distribution at NLO,
with a smaller effect on the $b \overline b$ separation than on
$c \overline c$.

However, introducing $k_T$ broadening has a significant effect on the azimuthal
correlations between heavy flavor hadrons, particularly for charm.  Introducing a relatively small $k_T$
broadening reduces the peak at $\phi = \pi$.  With $\Delta = 1$ in 
Eq.~(\ref{eq:avekt}), the peak at $\Delta \phi = \pi$
has disappeared and there is
instead an enhancement at $\Delta \phi = 0$.  This is because
$\langle k_T^2 \rangle^{1/2} \sim m_c$, causing the $\Delta \phi$ distribution
to become more isotropic,
losing its memory of the initial state kinematics.  The same effect happens to
a lesser extent for $b \overline b$ production because
$\langle k_T^2 \rangle^{1/2} < m_b$ so that the randomization that occurs for
$c \overline c$ production is decreased for the heavier quark.  Thus the
$\Delta \phi$ distribution can depend strongly on the value of $\Delta$ for
low $p_T$ charm quarks.  At higher heavy quark $p_T$, both the $c \overline c$
and $b \overline b$ azimuthal distributions have peaks at $\Delta \phi = \pi$
and 0.  These characteristics have been compared to reconstructed $D$ and $B$
mesons \cite{Vogt:2018oje} and to $b \overline b \rightarrow J/\psi J/\psi$
decays \cite{Vogt:2019xmm}.  Here we study how these kinematic characteristics might carry over into the heavy flavor decays to leptons.

\section{Decays of heavy flavors to dileptons}
\label{sec:decays}

The low mass dilepton contributions from charm and bottom decays are treated distinctly in these calculations because of their decays.  Only opposite sign dileptons arise from $c \overline c$ decays while both same sign and opposite sign dileptons can come from $b \overline b$ decays. Both charm and bottom quarks decay semileptonically with large branching ratios, on the order of 10\%.  An opposite sign lepton pair ($\mu^+ \mu^-$, $e^\pm \mu^\mp$, or $e^+e^-$) can result from correlated semileptonic decays of a $D \overline D$ meson pair.  The average branching ratio for $D \rightarrow lX$ is 9.6\%.

While a similar correlated decay is also possible from $B \overline B$ meson pairs, these decays can also produce like sign pairs.  For example, $B^- \rightarrow D^0 e^- \overline \nu_e$ while $B^+ \rightarrow \overline D^0 e^+ \nu_e$, producing an $e^+ e^-$ pair which can be detected.  It is also possible that, {\it e.g.}, the $B^- \rightarrow D^0 e^- \overline \nu_e$ decay is followed by a $D^0 \rightarrow K^- e^+ \nu_e$ decay.  This would result in an $e^+ e^-$ pair from a single decay with the first electron from the $B^- \rightarrow D^0 e^- \overline \nu_e$ (or $B^- \rightarrow e^-X$ below) denoted as a {\em primary} electron and the electron from the subsequent $D^0 \rightarrow K^- e^+ \nu_e$ (or $D^0 \rightarrow e^+ X$ below) referred to as a {\em secondary} electron.

The dileptons from $b \overline b$ decays calculated here include the like-sign subtraction.  It includes the following six contributions:
\begin{enumerate}
\item The opposite-sign lepton pair decay $B^+ \rightarrow e^+ X$, $B^- \rightarrow e^- X$ where the two primary electrons are detected;
\item The opposite-sign pair decay of $B^+ \rightarrow \overline D^0 e^+X$, $\overline D^0 \rightarrow e^- X$ and $B^- \rightarrow D^0 e^-X$, $D^0 \rightarrow e^+ X$ with the two secondary electrons forming the pair;
\item An opposite-sign pair from a single decay, $B^\pm \rightarrow D^0/\overline D^0 e^\pm$ followed by $D^0/\overline D^0 \rightarrow e^\mp X$ where the primary and secondary electrons form the pair;
\item An opposite-sign pair from a single decay, $B^\mp \rightarrow D^0/\overline D^0 e^\mp$ followed by $D^0/\overline D^0 \rightarrow e^\pm X$ where the primary and secondary electrons form the pair;
\item A like-sign pair with one electron from the primary decay $B^+ \rightarrow e^+ X$ and the other electron from the secondary decay $B^- \rightarrow D^0 e^- X$, $D^0 \rightarrow e^+ X$; 
\item A like-sign pair with one electron from the primary decay $B^- \rightarrow e^- X$ and the other electron from the secondary decay $B^+ \rightarrow \overline D^0 e^+ X$, $\overline D^0 \rightarrow e^- X$. 
\end{enumerate}
The four opposite-sign contributions are summed while the two like-sign contributions are subtracted.  While charged $B$ mesons are used as an example above, the decays of neutral $B$ mesons also contribute similarly, with the $\overline B^9$ taking the place of $B^+$ and $B^0$ replacing $B^-$ in the list of decay combinations above.

The same procedure can be followed for decays to muon pairs or $e \mu$ pairs.  The branching ratios are approximately the same for decays to electrons and muons with the branching ratio for the primary decay $\mathcal{B}(B \rightarrow eX) = 11$\% and that for the secondary decay $\mathcal{B}(B \rightarrow D \rightarrow eX) = 8.64$\%, assuming the branching ratios for the admixture of charged and neutral $B$ mesons \cite{ParticleDataGroup:2024cfk}.

\section{Dileptons from Heavy Flavor Decays at RHIC and the LHC}
\label{sec:calcs_for_expts}

Dileptons have been measured at both RHIC and the LHC.  Although all the collaborations at these colliders have the capability to measure lepton pairs, the focus in this work is on calculations in the acceptance of the PHENIX detector at RHIC and the ALICE detector at the LHC.  PHENIX had a broad lepton coverage, with the capability to measure both electrons and muons.  Likewise, the ALICE detector also measures electrons and muons.  However, there have been no similar studies of low mass continuum $\mu^+ \mu^-$ pairs.  Thus only $e^+ e^-$ pairs at central rapidity in the ALICE detector are considered here.  

The NLO calculations with the \textsc{HVQMNR} code outlined in the previous section are applied to study angular correlations between heavy flavor dileptons in $p+p$ collisions at both PHENIX (RHIC, $\sqrt{s} = 200$~GeV) and ALICE (LHC, $\sqrt{s} = 13$~TeV).

There are also other codes based on NLO calculations that have also be used to  study dileptons from heavy flavor decays.  \textsc{MC@NLO} \cite{Frixione:2003ei} and \textsc{POWHEG} ~\cite{Frixione:2007nw} employ event generators for hadronization by parton showers.  \textsc{MC@NLO} is a method for matching NLO QCD calculations to parton shower Monte Carlo simulations. Parton showers will generate terms that are
already present in the NLO calculations. To avoid double counting, the \textsc{MC@NLO} scheme removes such terms from the NLO expression. As a result, \textsc{MC@NLO} output contains events with negative weights.  \textsc{POWHEG} generates positive weight events only and can be interfaced to any shower Monte Carlo that is either $p_T$-ordered (e.g. \textsc{PYTHIA} \cite{Sjostrand:2007gs}), or allows the implementation
of a $p_T$ veto (e.g. \textsc{HERWIG++} \cite{Corcella:2002jc}), while avoiding any double counting when matching NLO calculations to a parton shower Monte Carlo.

Event generators like \textsc{PYTHIA} or \textsc{HERWIG} can also be used to directly simulate heavy flavor production.  However, since these event generators are leading order only, employing them forces categorization of individual diagrams according to event topology.  Indeed, production is characterized according to three different categories: flavor or pair creation, flavor excitation, and gluon splitting \cite{Bedjidian:2004gd}.  This classification ignores the fact that perturbative QCD calculations are sorted by the initial state: $gg$, $q \overline q$ and $g(q + \overline q)$.  The $gg$ contribution is dominant at collider energies over most of phase space, with $g(q + \overline q)$ the next largest contribution.  Thus flavor excitation and gluon splitting are real NLO contributions to heavy flavor production.

All three of the topologies treated separately by event generators are accessible through the $gg$ interaction while the $g(q + \overline q)$ falls under the domain of flavor excitation and $q \overline q$ is only involved in flavor creation.  The individual topologies might dominate production in certain kinematic regions but they cannot be treated as separate processes without mischaracterizing the cross section by ignoring interferences and virtual corrections.  Thus using such generators to characterize heavy flavor production through a certain diagram may be suggestive but incomplete.  Throughout this section, references will be made to how heavy flavor production topologies may appear in their correlated decays to epton pairs as arise from their production and decay in the \textsc{HVQMNR} code.   

\subsection{Calculations in the PHENIX Acceptance}
\label{sec:calcs_phenix}


PHENIX measured both electrons and muons and the angular correlations between them.  The muon spectrometers covered the forward ($1.2 < \eta < 2.4$) and backward ($-2.2 < \eta < -1.2$) pseudorapidity intervals, also referred to as the north and south arms.  Electrons were detected in the two central spectrometer arms (east and west) covering the pseudorapidity region $|\eta|<0.35$. Note the incomplete azimuthal coverage in the central region.  Because of this diverse coverage, PHENIX investigated the correlation of $Q \overline Q$ pairs through angular correlations of $e^+ e^-$, $e^\pm \mu^\mp$, and $\mu^+ \mu^-$ pairs~\cite{Aidala:2018dqb}.  The calculations here are in the PHENIX acceptance.  When calculating the decay electron contributions, the charged particle deflection in the azimuthal direction due to the magnetic field is taken into account, along with the radial location of the central detectors.  (See Ref.~\cite{Adare:2009qk} for details.)

The low invariant-mass region of $e^+ e^-$ or $\mu^+ \mu^-$ pairs is dominated by Dalitz and resonance decays of light-flavor hadrons. Hence, the measurement of dileptons from correlated open heavy-flavor hadron decays are typically limited to the intermediate invariant-mass region to efficiently suppress any light flavor contributions. However, this imposes strong kinematic constraints on the accessible $\Delta\phi$ region. The $e^\pm \mu^\mp$ pairs provide a great advantage as no mass selection is necessary to identify the signal. In addition, the rapidity gap between the arms of the detectors helps provides full coverage of $\Delta\phi$.

Muon pairs could be formed by both muons detected in either the north arm or the south arm as well as one muon in the south arm and the other in the north arm.  There are two possibilities for this latter case with {\it e.g.} the $\mu^+$ in the north arm and the $\mu^-$ in the south arm or the $\mu^-$ in the north arm and the $\mu^+$ in the north arm.  

PHENIX was able to separate the $c \overline c$ and $b \overline b$ contributions to $\mu^+ \mu^-$ pairs  by making a simultaneous fit to like sign and opposite sign pairs in dimuon mass and $p_T$.  The muon $p_T$ for both heavy flavors was $p_{T\mu} > 3$~GeV.  The $c \overline c$ mass range is $1.5 < m_{\mu \mu} < 2.5$ GeV while the $b \overline b$ mass range is $3.5 < m_{\mu \mu} < 10$ GeV.  

\begin{figure}[htp]
\centering
 \includegraphics[width=0.3\textwidth]{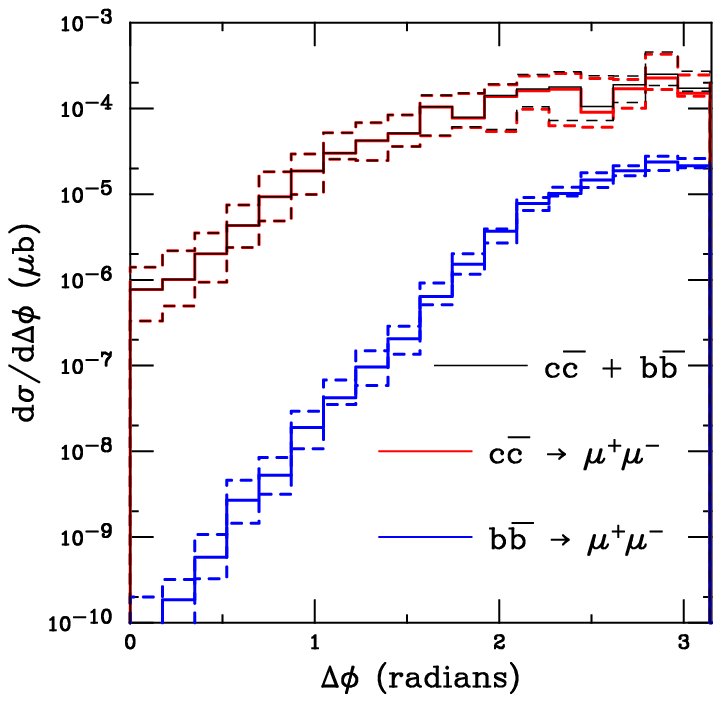}
 \caption{\label{fig:phenix_mumu}
   The azimuthal correlations (a) and mass distributions (b) for
   heavy flavor decays to $\mu^+ \mu^-$ pairs in the PHENIX acceptance are
   shown for $\Delta = 1$ in Eq.~(\ref{eq:avekt}).  The
   uncertainty bands for $c \overline c$ decays are given in red; those for
   $b \overline b$ decays in blue; while the sum of the two is given by the
   black histograms.  The central value is denoted by the solid histograms
   while the dashed histograms
   are the upper and lower limits of the uncertainty bands.
 }
\end{figure}

Figure~\ref{fig:phenix_mumu} shows the results for muon pairs. The muon pair azimuthal separation is expected to be peaked more back to back because of the muon acceptance.  The calculations present the $p+p$ cross sections as a function of $\Delta \phi$ as histograms representing the central value and upper and lower limits of the uncertainty bands for charm and bottom decays to lepton pairs.  The black histograms are the sum of the two results.  The total cross section from both decays is dominated by the charm decay contribution.

In Ref.~\cite{Aidala:2018dqb}, the muon pairs from $c \overline c$ decays are broader than those from $b \overline b$  which are more peaked at $\Delta \phi = \pi$.  This same behavior is observed in the calculations shown in Fig.~\ref{fig:phenix_mumu}(a).  As shown in Ref.~\cite{Vogt:2018oje}, when even rather minimal $k_T$ broadening is included, the $\Delta \phi$ distribution between the $c$ and $\overline c$ becomes more isotropic when integrated over all $p_T$.  However, when there is a finite minimum $p_T$, the pair correlation becomes more back to back once more.  Because the effect of broadening is smaller on the heavier bottom quarks, the corelation remains more strongly peaked back to back, even for a higher muon $p_T$.  It is worth noting that the lepton momentum cut of $p_\mu > 3$~GeV suggests that the initial quark momentum is considerably larger due to the nature of the three-body decay.

\begin{figure}[htp]
\centering
 \includegraphics[width=0.3\textwidth]{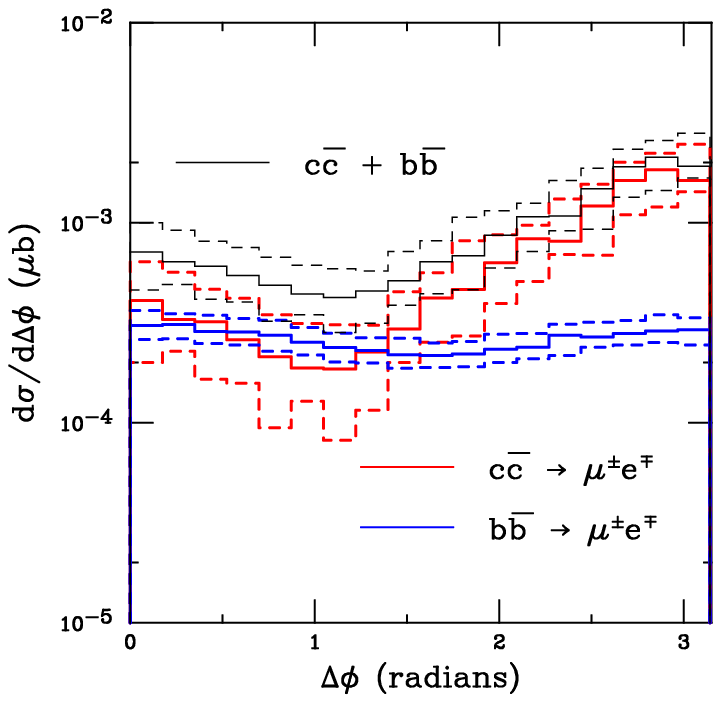}
 \caption{\label{fig:phenix_mue}
   The azimuthal correlations (a) and mass distributions (b) for
   heavy flavor decays to $e^\pm \mu^\mp$ pairs in the PHENIX acceptance are
   shown for $\Delta = 1$ in Eq.~(\ref{eq:avekt}).  The
   uncertainty bands for $c \overline c$ decays are given in red; those for
   $b \overline b$ decays in blue; while the sum of the two is given by the
   black histograms.  The central value is denoted by the solid histograms
   while the dashed histograms
   are the upper and lower limits of the uncertainty bands.
 }
\end{figure}

Figure~\ref{fig:phenix_mue} presents the results for electron-muon pairs.  There are eight possible combinations for $\mu e$ pairs: $\mu^\pm$ in the north arm and $e^\mp$ in the east or west arm; $\mu^\pm$ in the south arm and $e^\mp$ in the east or west arm; and the same configuration with opposite charges.

The momentum thresholds are lower in this analysis, with $p_{T,e} > 0.5$ GeV in the range $|\eta_e | < 0.35$ and $p_{T,\mu} > 1$ GeV in the range $1.4 < |\eta_\mu|< 2.1$.  The measurements \cite{PHENIX:2013klq} did not separate the results based on the flavor of the heavy meson initiating the decay.  In addition, unlike the $\mu^+ \mu^-$ and $e^+ e^-$ measurements, no mass range was specified, likely because there are no resonances that decay to different lepton flavors.  The results were more isotropic than those for muon pairs with the charm decay contribution being somewhat larger at $\Delta \phi \sim \pi$ and the two contributions becoming similar at $\Delta \phi \sim 0$.  The coverage of this combination, including both central rapidity for the electrons as well as forward and backward rapidity for the muons lends itself to more complete coverage of available phase space.

Reference~\cite{PHENIX:2013klq} showed results with large uncertainties for $p+p$ and d+Au colisions at $\sqrt{s} = 200$~GeV.  While the shape of the $p+p$ calculations matches the trend of the data well, the d+Au $\Delta \phi$ distribution is more isotropic, as one might expect from $k_T$ broadening in the nucleus \cite{Vogt:2018oje}.

In general, the uncertainty band for charm decays is broader than that for
bottom decays.  This is due to the smaller charm quark mass.  The upper limit of the charm cross section calculation is dominated by the mass uncertainty, even though it is small, $\pm 0.09$~GeV.  The mass uncertainty is larger for charm because, with $m_c = 1.27$~GeV, a change of $\pm 0.1$~GeV is 7\% of the total mass.  Also, decreasing $m_c$ to 1.18~GeV, effectively also reduced $\mu_F$ and $\mu_R$, especially at low $p_T$.  The lower limit of the charm band is dominated by the scale uncertainty, based on the minimum value of $\mu_F$, close to the minimum $Q^2$ of the parton densities.

On the other hand, the uncertainty band for bottom decays is narrower because of the larger quark mass so that the renormalization scale in the parton densities is always well above the minimum $Q^2$, even though the range in $\mu_F$ is smaller.  The mass uncertainty for $m_b = 4.65$~GeV is only 2\% of the quark mass so that even though the range of cross sections that make up the band is narrower for bottom than charm, the scale uncertainties dominate over the mass uncertainties for bottom.  The upper and lower limits of the bottom decay band arise from the lower and upper limits of the factorization scale range respectively.

Finally results are shown for $e^+ e^-$ pairs.  The pairs come from two mass regions, $1.15 < m_{ee} < 2.4$~GeV and $4.1 < m_{ee} < 8~$GeV. The electron $p_T$ is greater than 0.2~GeV in the measurement.  

\begin{figure}[htp]
\includegraphics[width=0.3\textwidth]{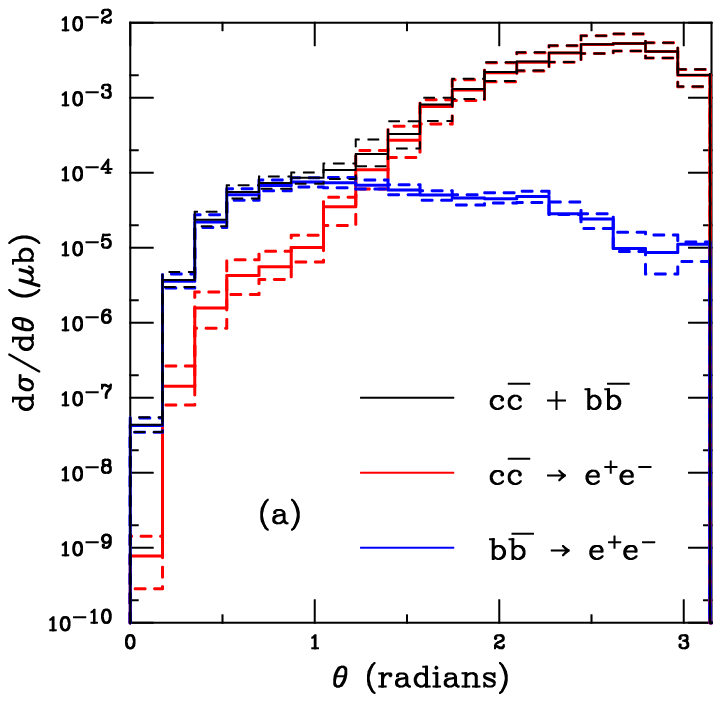}
\includegraphics[width=0.3\textwidth]{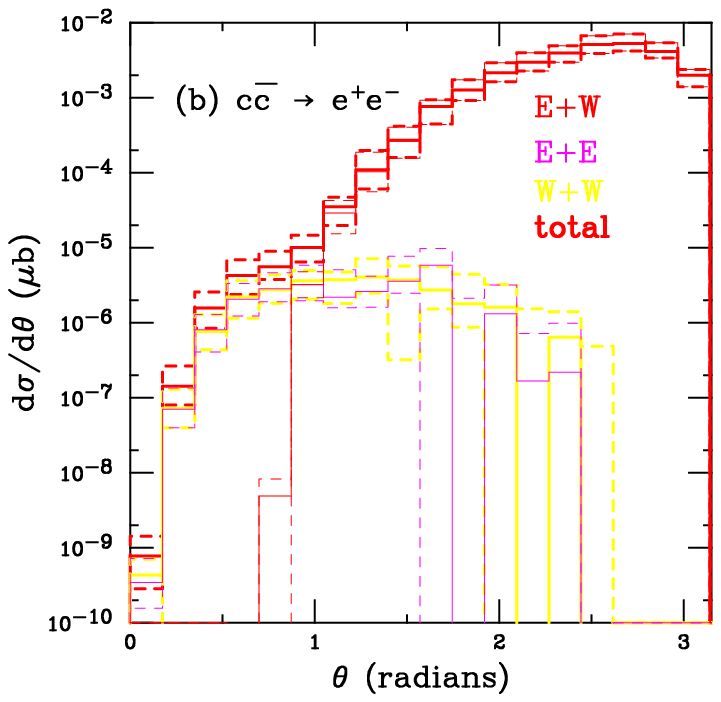}
\includegraphics[width=0.3\textwidth]{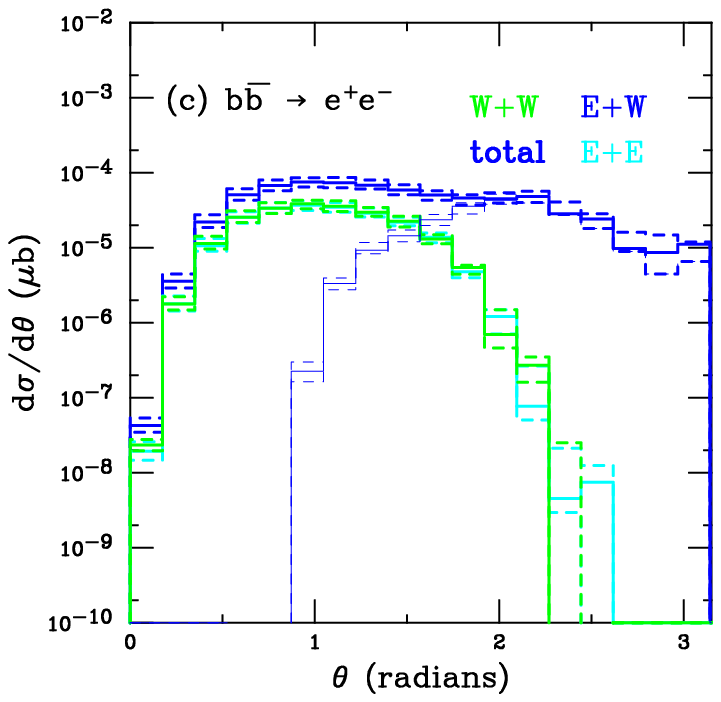}
\caption{\label{fig:phenix_epem}
  The azimuthal correlations (a-c) and mass distributions (d) for
  heavy flavor decays to $e^+ e^-$ pairs in the PHENIX acceptance are
   shown for $\Delta = 1$ in Eq.~(\ref{eq:avekt}).  (a) The
   uncertainty bands for $c \overline c$ decays are given in red; those for
   $b \overline b$ decays in blue; while the sum of the two is given by the
   black histograms.  The central value is denoted by the solid histograms
   while the dashed histograms
   are the upper and lower limits of the uncertainty bands.
   In (b) and (c) the charm and bottom contributions respectively are broken
   down according to which central
   spectrometer arms the dielectron was detected with pairs appearing in the
   opposite east and west arms (E+W) in the same color as the total and the
   other pairs appearing in the same arm (E+E and W+W) in different colors.
}
\end{figure}

There are also four possible combinations for forming electron pairs in the central arms, both the $e^+$ and $e^-$ in either the east arm or the west arm and one in each arm, $e^+$ in the east arm and $e^-$ in the west arm or the $e^-$ in the east arm and the $e^+$ in the west arm.  In this case the calculations are shown for all combinations in Fig.~\ref{fig:phenix_epem}(a) while the individual contributions from the possible contributions are shown separately for charm decays in (b) and bottom decays in (c).  The charm decay distribution is peaked for more back-to-back pairs while the bottom decay results are more isotropic in $\theta$ with a slight enhancement as $\theta \rightarrow 0$.

The reason for this behavior is apparent in Fig.~\ref{fig:phenix_epem}(b) and (c) where the calculations show the signal separated into different spectrometer arms.  When the two leptons of a pair are measured in a single arm, they are  forced to be in the range $\theta < \pi/2$.  The results are effectively independent of whether both the $e^+$ and the $e^-$ are detected in the west arm or the east arm.  When one lepton is detected in one arm and the opposite sign lepton in the other arm, the production is, by definition, near back to back.  In this case, there is a stronger contribution from charm decays.  The charm decay contribution is likely coming from the lower of the two mass ranges studied, $1.1 < m_{ee} < 2.4$~GeV, where charm is dominant, while the bottom decay contribution is stronger in the higher mass range covered.

The PHENIX $e^+ e^-$ pair results were first published in Ref.~\cite{PHENIX:2017ztp}.  In that work, the $e^+ e^-$ results were also shown in two different electron $p_T$ ranges, $p_T < 3$~GeV and $3 < p_T < 8$~GeV.  Although the same $p_T$ separation is not shown here, it is clear from Ref.~\cite{Vogt:2018oje} that the charm signal dominates at lower $p_T$ because the average charm $p_T$ is smaller compared to the bottom quark $p_T$ due to the larger bottom mass, $\langle p_{T\, c} \rangle \sim 1.7$~GeV compared to $\langle p_{T \, b} \rangle \sim 5$~GeV \cite{Vogt:2018oje}.  The higher the quark $p_T$, the more likely it is that there will be a significant contribution at $\theta = 0$.  Thus the $b$ quark contribution shown here is predominantly from the high $p_T$ part of the electron spectrum.

PHENIX reported the total $c \overline c$ and $b \overline b$ cross sections in Ref.~\cite{Aidala:2018dqb} based on \textsc{PYTHIA} and \textsc{POWHEG}.  
The \textsc{PYTHIA} results  were tuned to the PHENIX $e^+ e^-$ invariant mass distribution \cite{Adare:2008ac}.  The cross sections from \textsc{POWHEG} are $\sigma_{c \overline c} = 318$~$\mu$b and $\sigma_{b \overline b} = 3.94$~$\mu$b.  The corresponding cross sections from \textsc{PYTHIA} are $\sigma_{c \overline c} = 348$~$\mu$b and $\sigma_{b \overline b} = 3.59$~$\mu$b.
No uncertainties on these values were presented.  Both of these calculations used $m_c = 1.29$~GeV and $m_b = 4.1$~GeV with scale factors $\mu^2 = p_T^2 + m^2$.

The \textsc{HVQMNR} code gives $\sigma_{c \overline c} = 459.5^{+179}_{-131.8}$~$\mu$b and $\sigma_{b \overline b} = 2.09^{+0.37}_{-0.31}$~$\mu$b, after having been fit to the available total $c \overline c$ and $b \overline b$ cross section data, as described in Sec.~\ref{sec:model}.  These central values are about 40\% above the simulated PHENIX values given for $c \overline c$ and about 76\% below those given for $b \overline b$.  The charm quark mass used in these NLO calculations are $m_c = 1.27$~GeV and $m_b = 4.65$~GeV while the scales are proportional to the transverse mass of the $Q \overline Q$ pair, $m_T^2 = m^2 + 0.5(p_{T_Q}^2 + p_{T_{\overline Q}}^2)$.  In the case of charm quarks, even though the masses are very similar in these calculations and the PHENIX simulations, the larger factorization scale used here will result in a larger central cross section.  Given the uncertainty on the mass as well as the relatively large scale range for charm production given in Sec.~\ref{sec:model}, the cross section calculated here is in agreement with that obtained by PHENIX.  On the other hand, the bottom quark mass used in the PHENIX simulations is 13\% smaller than that used in these calculations so that one may expect a smaller total cross section in this calculation.  The narrow scale uncertainty results in smaller uncertainty band than obtained for charm production.  The relative differences in these cross sections can likely account for any differences in the relative shapes of these calculations in comparison to the data.

\subsection{Predictions for ALICE at 13~TeV}
\label{sec:calcs_alice}

While the ALICE collaboration measures electrons in the central rapidity region and muons in ther forward muon detector, they have mainly concentrated on measurements of electron pairs.  For an example of low mass muon pair measurements, see Ref.~\cite{ALICE:2011ad}, concentrating on low mass vector meson yields.  No analyses of $e$-$\mu$ coincidences have yet been made available.

\begin{figure}[htp]
\includegraphics[width=0.3\textwidth]{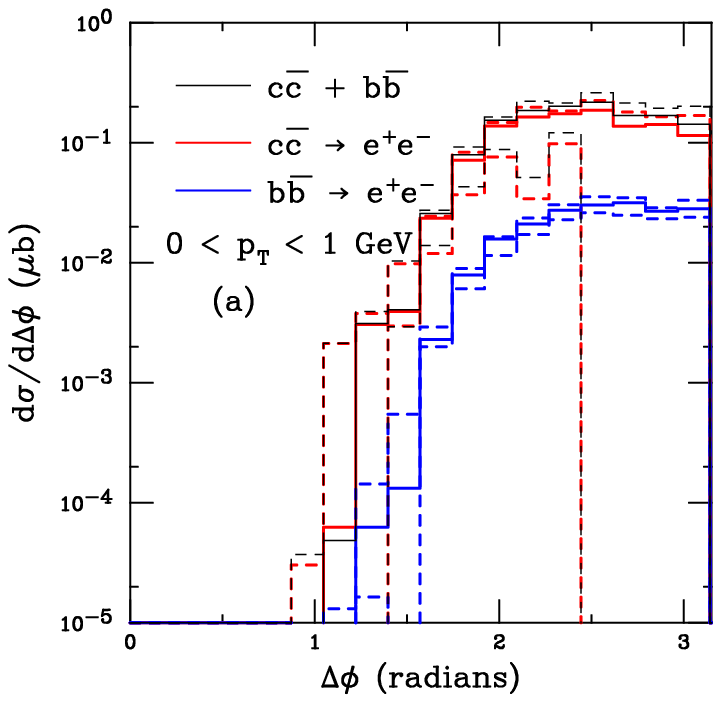}
\includegraphics[width=0.3\textwidth]{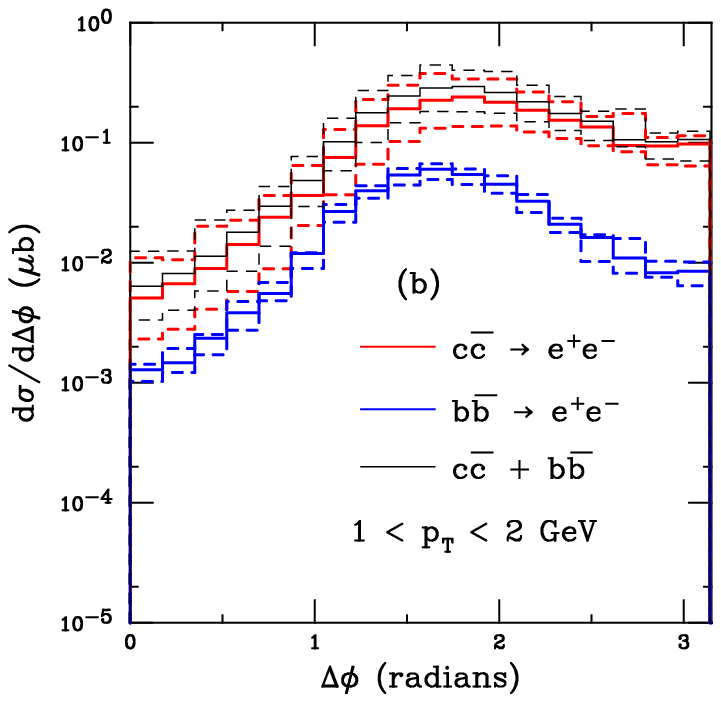}
\includegraphics[width=0.3\textwidth]{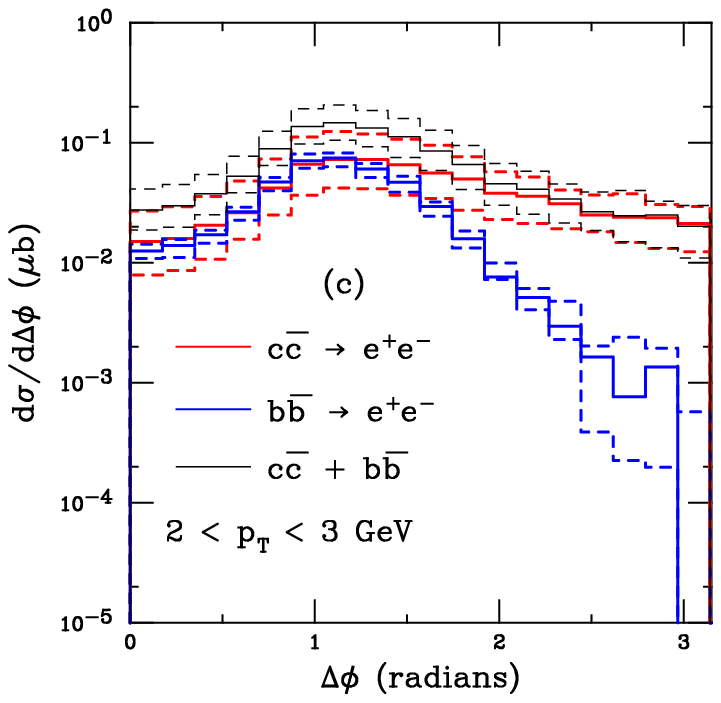}\\
\includegraphics[width=0.3\textwidth]{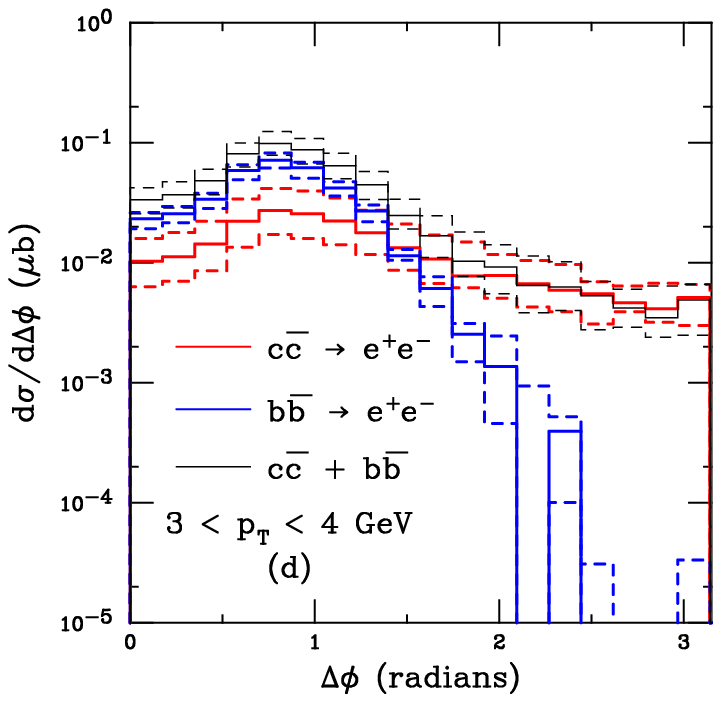}
\includegraphics[width=0.3\textwidth]{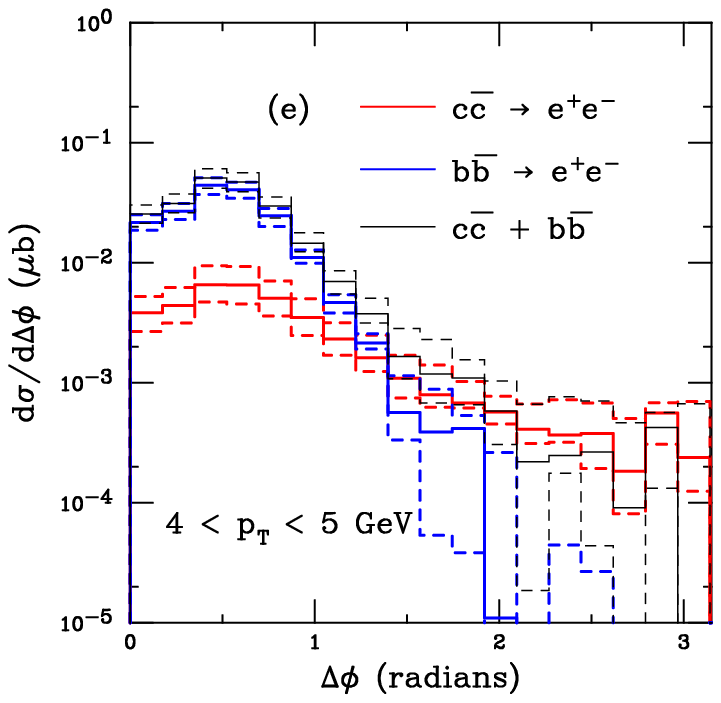}
\includegraphics[width=0.3\textwidth]{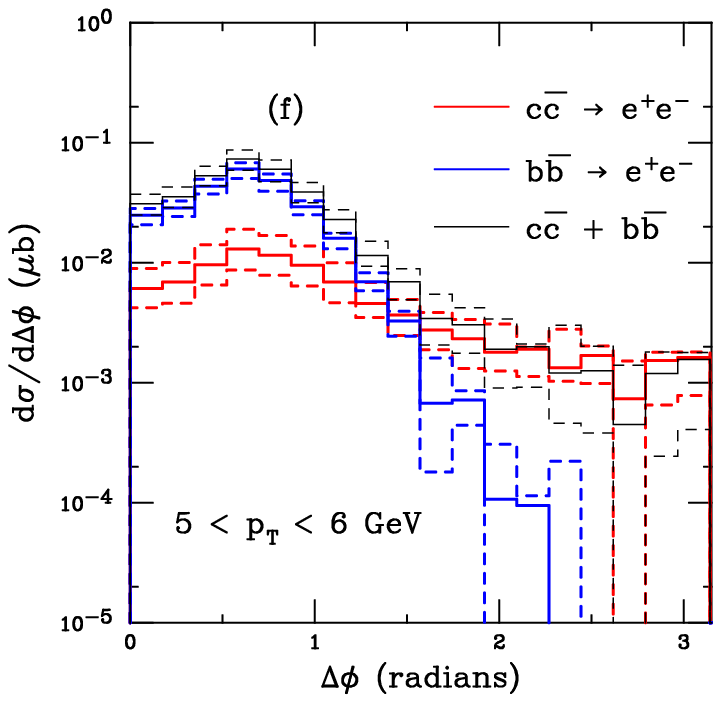}
\caption{\label{fig:alice_epem}
  The azimuthal correlation between
  heavy flavor decays to $e^+ e^-$ pairs in mass region $1.1<M_{ee}<2.8$~GeV
  at $\sqrt{s} = 13$~TeV in the ALICE acceptance are
   shown for $\Delta = 1$ in Eq.~(\ref{eq:avekt}).  The
   uncertainty bands for $c \overline c$ decays are given in red; those for
   $b \overline b$ decays in blue; while the sum of the two is given by the
   black histograms.  The central value is denoted by the solid histograms
   while the dashed histograms
   are the upper and lower limits of the uncertainty bands.
   Results are given for different $p_T$ bins: (a) $p_T < 1$~GeV,
   (b) $1 < p_T < 2$~GeV, (c) $2 < p_T < 3$~GeV, (d) $3 < p_T < 4$~GeV,
   (e) $4 < p_T < 5$~GeV, and (f) $5 < p_T < 6$~GeV.
}
\end{figure}

In Ref.~\cite{ALICE:2018gev}, the ALICE collaboration studied the mass and $p_T$ distributions of $e^+ e^-$ pairs in the central rapidity region for electron $p_T > 0.2$ GeV and $|\eta_e| < 0.8$.  They binned the results into four electron pair $p_T$ regions:   $p_T < 1$~GeV, $1 < p_T < 2$~GeV, $2 < p_T < 3$~GeV and $3 < p_T < 6$~GeV.  They did not present azimuthal correlations.  Here the results, calculated these correlations in the ranges: $p_T< 1$~GeV, $1 < p_T < 2$~GeV, $2 < p_T< 3$~GeV, $3 < p_T < 4$~GeV, $4 < p_T< 5$~GeV and $5 < p_T < 6$~GeV are presented in Fig.~\ref{fig:alice_epem}.  The correlations are given in the mass range $1.1 < m_{ee} < 2.7$~GeV in $p+p$ collisions at $\sqrt{s} = 13$~TeV. The simultaneous selection in mass and $p_T$ put strong constraints on the accessible phase space in $\Delta\phi$, as shown in Fig.~\ref{fig:alice_epem}.

In the lowest $p_T$ bin, there are no pairs produced at $\Delta \phi \sim 0$ and the charm decay contribution dominates.  As the electron pair $p_T$ increases, the peak in the $\Delta \phi$ distribution shifts from $\Delta \phi \sim \pi$ to $\Delta \phi$ approaching 0.  The dominant contribution also shifts from charm decays to bottom decays, particularly for low $\Delta \phi$.  While some signal from charm decays remains at $\Delta \phi \sim \pi$, the statistical significance become smaller as the charm decay electrons come from increasingly higher $p_T$ charm hadrons.  Conversely, while the $b$ decay contributions grow at small $\Delta \phi$, they are absent at larger $\Delta \phi$ for electrons with pair $p_T$ greater than 3~GeV.  This behavior can be attributed to the pair mass constraint here.  The results are also consistent with those shown in the more constrained $e^+ e^-$ results from PHENIX in Fig.~\ref{fig:phenix_epem}.

\begin{figure}[htp]
\includegraphics[width=0.3\textwidth]{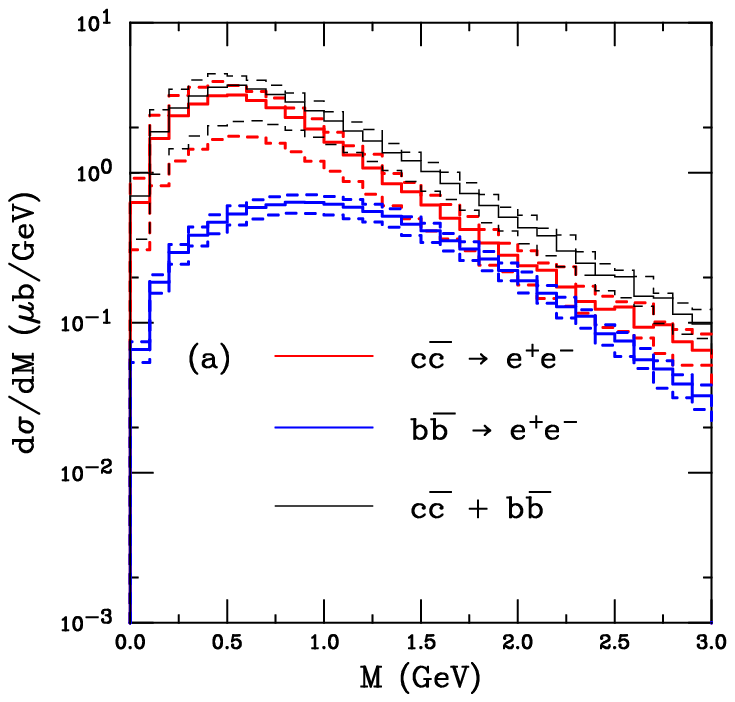}
\includegraphics[width=0.3\textwidth]{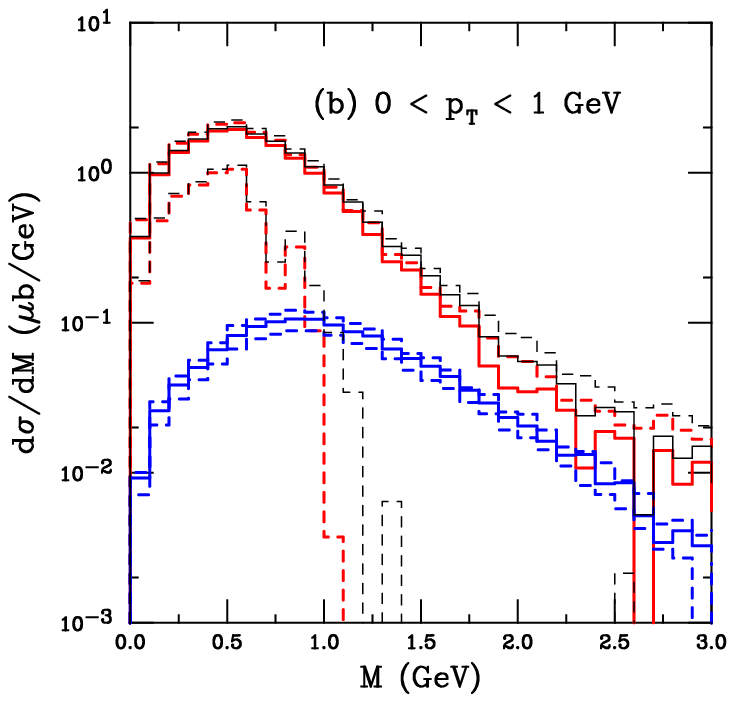}
\includegraphics[width=0.3\textwidth]{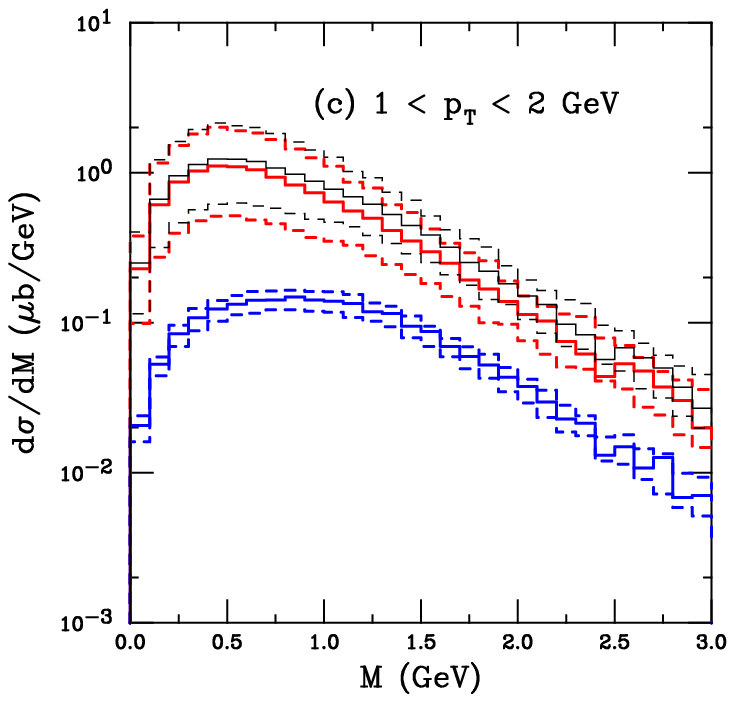}\\
\includegraphics[width=0.3\textwidth]{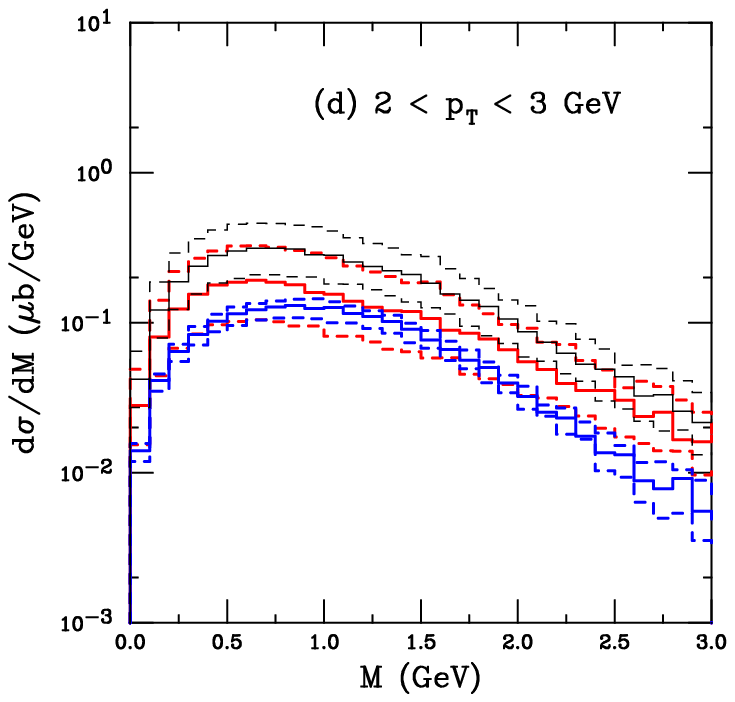}
\includegraphics[width=0.3\textwidth]{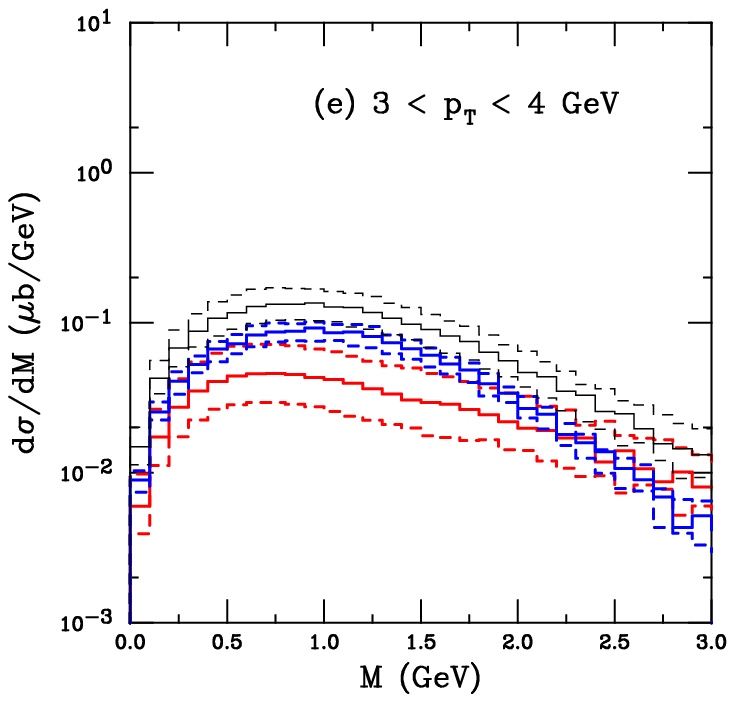}
\includegraphics[width=0.3\textwidth]{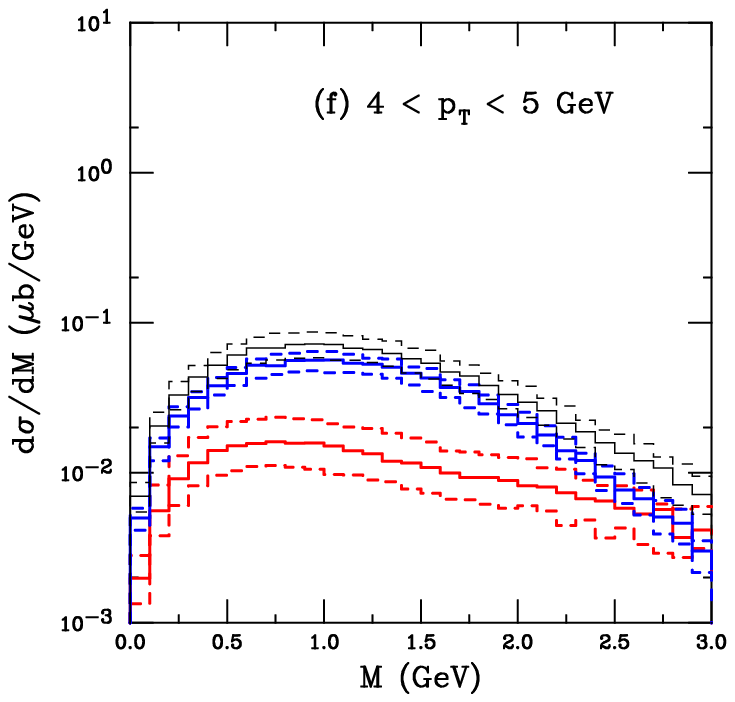}
\caption{\label{fig:alice_mass}
  The mass distributions of
  heavy flavor decays to $e^+ e^-$ pairs
  at $\sqrt{s} = 13$~TeV in the ALICE acceptance are
   shown for $\Delta = 1$ in Eq.~(\ref{eq:avekt}).  The
   uncertainty bands for $c \overline c$ decays are given in red; those for
   $b \overline b$ decays in blue; while the sum of the two is given by the
   black histograms.  The central value is denoted by the solid histograms
   while the dashed are the upper and lower limits of the uncertainty bands.
   Results are given for different $p_T$ bins: (a) all $p_T$,
   (b) $p_T < 1$~GeV, (c) $1 < p_T < 2$~GeV, (d) $2 < p_T < 3$~GeV,
   (e) $3 < p_T < 4$~GeV, and (f) $4 < p_T < 5$~GeV.
}
\end{figure}

The mass distributions are presented in Fig.~\ref{fig:alice_mass}.  Here the mass distribution integrated over all pair $p_T$ are shown in Fig.~\ref{fig:alice_mass}(a) while the five subsequent subfigures show results for pair $p_T$ values of $0< p_T < 1$~GeV (b), $1 < p_T < 2$~GeV, $2< p_T <3$~GeV, $3 < p_T < 4$~GeV and $4< p_T < 5$~GeV.  The mass distributions are shown up to 3~GeV where the distinction between charm and bottom decays is clearest.  Above this mass, the calculational statistics become poor.  This region also more closely corresponds to the low mass region covered by the calculations in Fig.~\ref{fig:alice_epem}.  Similar to those results, one can see that charm decays dominate the mass distribution for low electron pair $p_T$ but as the $p_T$ increases, the bottom decay contributions clearly dominate.

The ALICE collaboration reported the $c \overline c$ and $b \overline b$ cross sections at $y = 0$ in the mass range $1.03 \leq m_{ee} \leq 2.86$~GeV.  The inidividual $Q \overline Q$ cross sections were determined by fitting the $e^+ e^-$ data simultaneously, including all contributions to lepton pairs in the invariant mass range and $p_T < 6$~GeV using the $c \overline c$ and $b \overline b$ production and decay templates from \textsc{PYTHIA} and \textsc{POWHEG}.  The contributions from light hadron and $J/\psi$ decays were kept fixed while the normalizations for charm and bottom were allowed to vary.

While the two generators give different shapes, they both give good fits to the ALICE data~\cite{ALICE:2018gev}.  The shapes differ in the two simulations because the $Q \overline Q$ correlations differ.  As discussed at the beginning of this section, codes like \textsc{PYTHIA} are leading order only so that $Q \overline Q$ production is based on event topology rather than on initial state.  Because, for example, the $gg$ initial state contains contributions from all three different event topologies: flavor creation, flavor excitation and gluon splitting, results from \textsc{PYTHIA} have different weights for certain diagrams resulting from the same initial state, correlations between the $Q$ and $\overline Q$ will differ, leading to differences in the lepton pair distributions.  However, even though the $Q \overline Q$ shapes vary in \textsc{PYTHIA} and \textsc{POWHEG}, they do so in such a way that the agreement of their sum remains good in both approaches.

In Ref.~\cite{ALICE:2018gev}, the charm cross section extracted in the intermediate mass region at $y=0$ were $974 \pm 138 ({\rm stat.}) \pm 140 ({\rm syst.})$~$\mu$b for \textsc{PYTHIA} and $1417 \pm 184 ({\rm stat.}) \pm 204 ({\rm syst.})$~$\mu$b for \textsc{POWHEG}, based on the same branching ratio for $D \rightarrow e X$ used here, 9.6\%.  In the same mass interval, the $p_T$-integrated distribution shown in Fig.~\ref{fig:alice_mass} from the \textsc{HVQMNR} code at $y=0$ is $611^{+111}_{-189}$~$\mu$b, lower than both of the ALICE cross sections but within $1\sigma$ of the \textsc{PYTHIA} result.  In later work, the ALICE Collaboration revised their charm cross sections upward by a factor of 1.72 based on a lower total branching ratio of charm hadrons to electrons, $7.33^{+0.61}_{-0.84}$\% instead of 9.6\% \cite{Addendum:ALICE}, finding $1674 \pm 237 ({\rm stat.}) \pm 241 ({\rm syst.}) \, ^{+387}_{-283} ({\rm BR)}$~$\mu$b for \textsc{PYTHIA} and $2435 \pm 316 ({\rm stat.}) \pm 351 ({\rm syst.}) \, ^{562}_{-411} ({\rm BR})$~$\mu$b for \textsc{POWHEG}.  Here the additional uncertainty is due to the branching ratio to leptons \cite{Addendum:ALICE}. Note that the template shapes for the charm and bottom cross sections were not changed, only the branching ratios.  Adjusting the branching ratio in these calculation to that of Ref.~\cite{Addendum:ALICE} would increase the \textsc{HVQMNR} cross section in the same range to $1048^{+190}_{-324}$~$\mu$b, keeping the level of agreement similar.

The $b \overline b$ cross sections reported in Ref.~\cite{ALICE:2018gev} were unaffected by the adjustment of the charm branching ratio in Ref.~\cite{Addendum:ALICE}.  They were reported as $79 \pm 14 ({\rm stat.}) \pm 11 ({\rm syst.})$~$\mu$b for \textsc{PYTHIA} and $48 \pm 14 ({\rm stat.}) \pm 7 ({\rm syst.})$~$\mu$b for \textsc{POWHEG}, based on a branching ratio of bottom hadrons to electrons of 21.53\%, a factor of 2.32 times larger than the approximately 10\% branching ratio used here.  (This 10\% average is based on an average of the 11\% for primary $B$ decays and 8.64\% for secondary $B$ decays described in Sec.~\ref{sec:decays}.  Because the like-sign subtraction is carried out within the same execution of the \textsc{HVQMNR} code, with all 6 decay possibilities included in a single instance, it is not possible to separate primary-primary, primary-secondary, and secondary-secondary decays from one another to provide a more exact branching ratio for $e^+e^-$ pairs from $b \overline b$ decays in this calculation.)  Using this $\sim 10$\% branching ratio, the \textsc{HVQMNR} cross section in the intermediate mass region at $y=0$ and integrated over $p_T < 6$~GeV is $323^{+40}_{-51}$~$\mu$b.  Dividing this value by the square of the difference in branching ratios gives a $b \overline b$ cross section of $69^{+8.6}_{-11}$~$\mu$b here, in good agreement with the ones extracted by ALICE.  

\section{Sensitivity of the results to $\langle k_T^2 \rangle$}
\label{sec:Delta_kt}

\begin{figure}[htp]
\centering
 \includegraphics[width=0.3\textwidth]{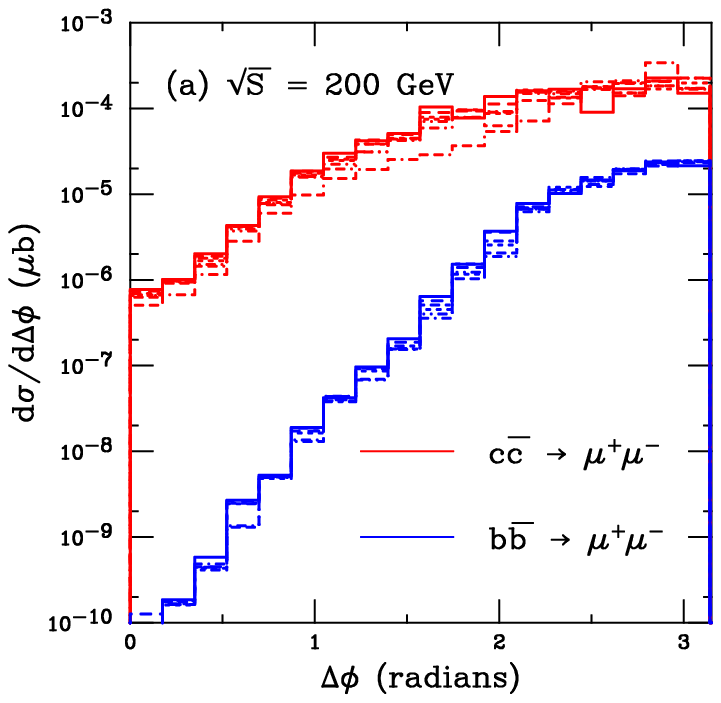}
 \includegraphics[width=0.3\textwidth]{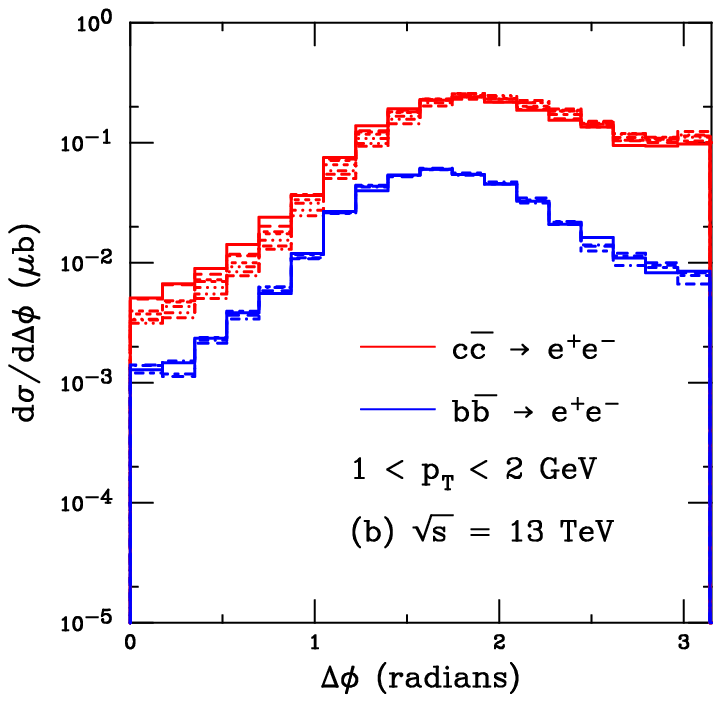}
\caption{\label{fig:phenix_mumu_Delta} Variation of the angular correlation between muon pairs for PHENIX (a) and electron pairs with $1 < p_T < 2$~GeV for ALICE (b) due to changing the parameter $\Delta$ in Eq.~(\ref{eq:avekt}).  The solid curve is the default value, $\Delta = 1$, with the values of $1/2$ (dashed), 0 (dot-dashed), $-1/2$ (dotted), $-1$ (dot-dot-dot-dashed), $-3/2$ (dot-dot-dash-dashed), and $\langle k_T^2 \rangle = 0$ (dot-dash-dash-dashed). Note that in this case, $\Delta = -3/2$ and $-1$ are not allowed for $b \overline b$ because it results in a negative $\langle k_T^2 \rangle$.}
\end{figure}

The change in the results that arise from varying $\Delta$ are shown in Fig.~\ref{fig:phenix_mumu_Delta} for $\mu^+ \mu^-$ pairs from PHENIX and $e^+e^-$ pairs with $1 < p_T < 2$~GeV for ALICE.  These particular results were chosen because they span the entire $\Delta \phi$ range and were of high statistical accuracy, limiting fluctuations.  The average values of $\langle k_T^2 \rangle$ for $c \overline c$ production are 1.19~GeV$^2$ at $\sqrt{s} = 200$~GeV and 1.54~GeV$^2$ at 13~TeV while those for $b \overline b$ are 1.76~GeV$^2$ at 200~GeV and 3.16~GeV$^2$ at 13~TeV.  The curves for charm include $\langle k_T^2 \rangle = 0$ and $\Delta = -3/2, -1, -1/2, 0, 1/2$ and 1 while those for bottom include  $\langle k_T^2 \rangle = 0$ and $\Delta = -1/2, 0, 1/2$ and 1.  There are fewer curves for bottom quarks because the value of $n=3$ for bottom means that $\Delta = -3/2$ and $-1$ would give negative values of $\langle k_T^2 \rangle$.

Results for the variation of $\Delta$ were shown for $c \overline c$ and $b \overline b$ quark pairs in Ref.~\cite{Vogt:2018oje} for central rapidities at $p_T < 10$~GeV and $p_T > 10$~GeV at $\sqrt{s} = 7$~TeV.  There was significant variation of the $\phi$ distribution at low $p_T$, especially for charm production.  Without any $k_T$ kick, there was a strong peak close to $\phi = \pi$, with a long tail toward lower $\phi$.  Adding a $k_T$ kick made the $\phi$ distribution more isotropic, especially for charm quarks because the lower mass meant that a $k_T$ kick of 1~GeV$^2$ or so is similar to the magnitude of the charm quark mass, $\langle k_T^2 \rangle \sim m_c^2$, making it relatively easy to change the direction of motion of the charm quark.  On the other hand, the heavier $b$ quark mass makes it relatively more immune to even a rather large $k_T$ kick.  Increasing the $k_T$ kick changes the $\phi$ distributions from back-to-back peaked to more isotropic, with an enhancement at $0^\circ$ for charm quark pairs.  At high $p_T$ both the charm and bottom quark pair $\phi$ distributions are independent of the size of the $k_T$ kick since $m_T > \langle k_T^2 \rangle^{1/2}$.  Here there are two distinct peaks, one at $\phi \sim \pi$ for back-to-back production and at $\phi = 0$ for $Q \overline Q$ pairs produced opposite a high $p_T$ light parton.

These distinct characteristics can be modified when heavy quark decays to leptons are considered.  The charm and bottom quark semileptonic decays are assumed to be isotropic in the rest frame of the parent heavy quark albeit boosted in the heavy quark direction of travel in the lab frame.  Thus a strong $k_T$ kick dependence in the initially-produced heavy quarks may be absent when considering their decay leptons.  However, it is worth noting that the variation in the results due to $k_T$ broadening is similar to those due to the mass and scale parameters.

\section{Summary}

Azimuthal correlations between lepton pairs from semileptonic decays of charm and bottom pairs were studied at next-to-leading order, including fragmentation and $k_T$ broadening.  While the effects of $k_T$ broadening can be significant at low heavy quark $p_T$ \cite{Vogt:2018oje}, the isotropization introduced by the decays reduces the sensitivity to the broadening.  This decorrelation effect was also seen for $b \overline b \rightarrow J/\psi J/\psi$ decays \cite{Vogt:2019xmm}.

In $p+p$ collisions the broad characteristics of heavy flavor production are retained in the lepton pair distributions.  Low $p_T$ leptons, generally from lower $p_T$ heavy flavors, tend to be produced back to back, while at higher $p_T$, the lepton pairs are more likely to come from production of a heavy quark pair with an energetic light parton, resulting in smaller angular separation, as seen in Fig.~\ref{fig:alice_epem}.  However, enhanced broadening in $p+A$/d+Au and $A+A$ collisions, would further reduce most of the kinematic characteristics retained from the production, particularly at low $p_T$ \cite{Vogt:2019xmm}.  To study the angular distribution from non-prompt heavy-flavor decay dileptons in $A+A$ collisions, the prompt thermal dilepton contribution would first have to be removed, based on impact parameter or decay length, or focus on $e \mu$ pairs that cannot be produced electromagnetically.  

{\bf Acknowledgements:}
This work was performed under the auspices of the U.S. Department of Energy by Lawrence Livermore National Laboratory under  Contract DE-AC52-07NA27344 and supported by the U.S. Department of Energy,  Office of Science, Office of Nuclear Physics (Nuclear Theory) under contract number DE-SC-0004014. This research was supported by the DFG cluster of excellence ``Origin and Structure of the Universe''. This research was supported by the Munich Institute for Astro- and Particle Physics (MIAPP) of the DFG cluster of excellence ``Origin and Structure of the Universe''.  RV thanks the Technical University of Munich for hospitality at the beginning of this work.

\bibliography{bibliography}


\end{document}